\documentstyle [epsf]{article}
\begin{document}
\title{Does femtosecond time-resolved second-harmonic generation probe electron temperatures at surfaces?}
\author{J. Hohlfeld, U. Conrad,  and E. Matthias\\
Fachbereich Physik, Freie Universit\"at Berlin, Arnimallee 14,
 14195 Berlin,\\ Germany}

\date{ }
\maketitle
\noindent {\bf Abstract}
  Femtosecond pump-probe second-harmonic generation (SHG) and
  transient linear reflectivity measurements were carried out on
  polycrystalline Cu, Ag and Au in air to analyze whether the electron
  temperature affects Fresnel factors or nonlinear susceptibilities,
  or both. Sensitivity to electron temperatures was attained by using
  photon energies near the interband transition threshold. We find
  that the nonlinear susceptibility carries the electron temperature
  dependence in case of Ag and Au, while for Cu the dependence is in
  the Fresnel factors. This contrasting behavior emphasizes that SHG
  is not {\it a priori} sensitive to electron dynamics at surfaces or
  interfaces, notwithstanding its cause.\\
\noindent {\bf PACS:}  42.65.Ky, 78.47.+p, 63.20.Kr\\
\\
\\
Nonlinear optical techniques like second-harmonic and sum-frequency
generation gain increasing importance for the investigation of
surfaces \cite{She_94}, interfaces \cite{Ric_88}, thin films
\cite{Aus_95}, and multilayers \cite{Ras_94}. This trend is
intensified by the possibility to investigate the electron dynamics
with femtosecond time resolution. For example, Hicks {\it et al.}\
\cite{Hic_88} used second-harmonic generation on Ag(110)
surfaces to study the time dependence of lattice temperature after
pulsed laser excitation, and ultrafast laser-induced order/disorder
transitions were detected by pump-probe SHG in semiconductors
\cite{Tom_88,Sae_91,Sok_95}. A broader application of these ultrafast
techniques to surface and thin film physics is expected, but increased
use of time-resolved SHG requires a more detailed understanding of the
influence of transient electron temperatures on the nonlinear signal
\cite{Hic_88,Hoh_95}.

The SHG yield is determined by the linear optical properties of the material
for the fundamental and frequency-doubled radiation, given by the
Fresnel factors $f(\omega)$ and $F(2\omega)$, and its intrinsic
ability to generate the second harmonic, characterized by the
nonlinear susceptibility, $\chi^{(2)}$ \cite{Sip_87}.  All three
quantities may be affected by the electron temperature. While
$f(\omega)$ and $F(2\omega)$ sample the material within their
optical penetration depth, $\chi^{(2)}$ is surface sensitive
within the dipole approximation for centrosymmetric materials. Hence,
in order to uncover possible differences in electron dynamics of bulk
material, surfaces, and thin films, it is of importance to isolate the
temperature dependence of each individual factor.

In this contribution, femtosecond time-resolved SHG in combination with
transient linear reflection at frequencies $\omega$ and $2\omega$ was
utilized to study the temperature dependence of $\chi^{(2)}$ for
polycrystalline Cu, Ag, and Au samples. It is demonstrated that
pump-probe SHG is particularly sensitive to transient electron
temperatures when photon energies near interband transitions are used.
The $2\,$eV photon energy of the laser employed nearly matches the
energy difference between d-band and Fermi energy for Cu and Au
\cite{Ees_83,Sch_87}, while in case of Ag this
separation is about $3.9\,$eV \cite{Chr_72} and thus almost resonant with
$2\hbar\omega$. Broadening and shifting of the Fermi distribution due
to absorption leads to changes in electronic occupancy which
sensitively affects both the dielectric function \cite{Hoh_95,Sch_87}
and the nonlinear susceptibility \cite{Kel_86}. The influence of
lattice temperature increase on optical properties is negligible for
our experimental conditions \cite{Ees_83}.

\section{Experimental details}

\begin{figure}[htbp]
\epsfxsize=12.cm
\centerline{\epsffile{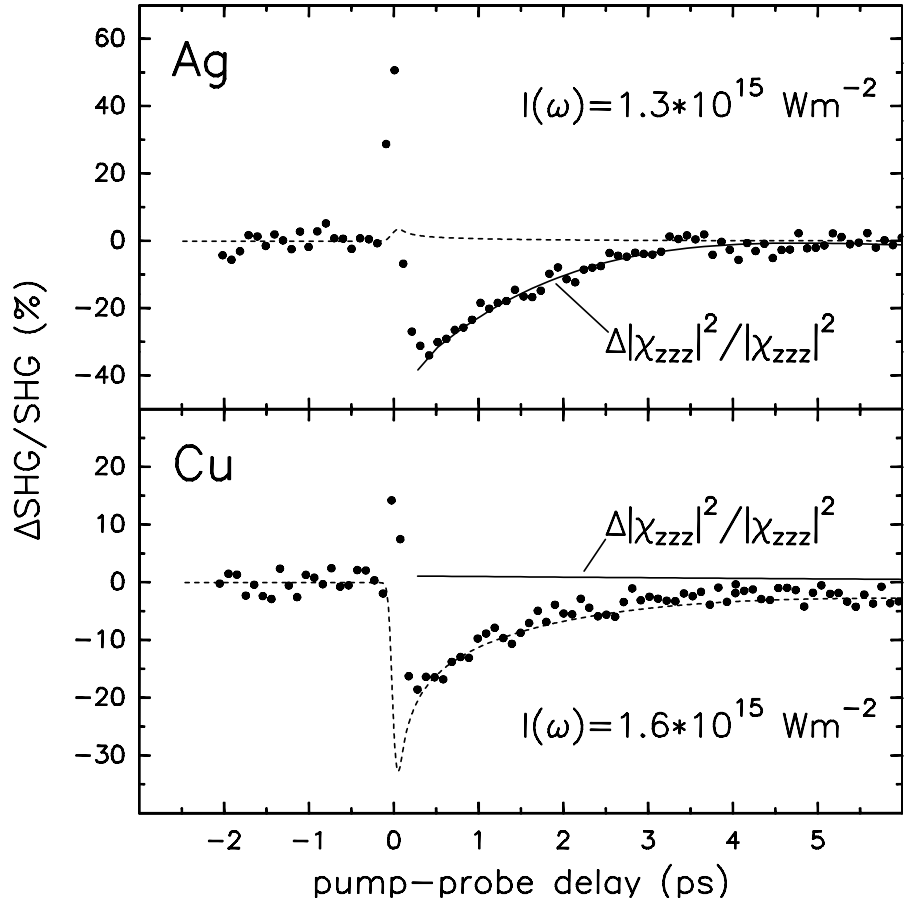}}
\vspace{2ex}
\caption{\label{2}
  Relative changes of SHG yield (dots) versus pump-probe delay time
  for Ag and Cu. The data are normalized to the SHG yield of the probe
  pulses at negative delays. Dotted lines indicate the electron
  temperature dependence of SHG due to the Fresnel factors only. Solid
  lines originate from the difference between smoothed experimental
  data and dotted lines and represent the relative change of
  $|\chi_{zzz}|^{2}$ with electron temperature.}
\end{figure}

\begin{figure}[htbp]
\epsfxsize=12.cm
\centerline{\epsffile{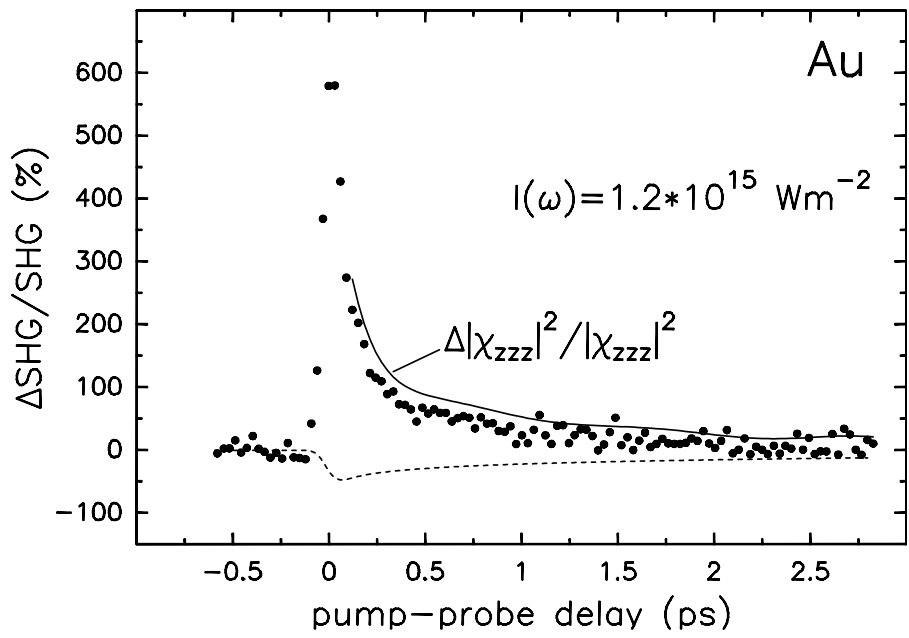}}
\vspace{2ex}
\caption{\label{3}
  Relative changes of SHG yield (dots) versus pump-probe delay time
  for the $2.5\,\mu$m Au sample. Description of solid and dotted lines
  as in Fig.\ \protect{\ref{3}}.}
\end{figure}

The experimental setup was reported in detail in \cite{Hoh_95}. Laser
pulses of $100\,$fs duration at a wavelength of $630\,$nm ($2\,$eV)
were generated by a colliding-pulse mode locked dye laser and
amplified to $50\,\mu$J at repetition rates of a few Hz.  For all SHG
measurements, identical pump and probe pulses of energies between $12$
and $16\,\mu$J and foci of $0.2\,$mm were used, and the change of the
probe pulse SHG yield due to the pump pulse was recorded.  The choice
of identical pump and probe pulse energies caused strong heating due
to the probe pulse, but was necessary to obtain detectable SHG yield.
The effective area monitored by SHG corresponded to about one half of
the beam area and was therefore nearly homogeneously heated. Great
care was taken to work at intensities that caused no surface damage.
Probe pulses were delayed with respect to pump pulses by a computer
controlled delay stage with $0.1\,\mu$m accuracy and a typical step
size of $3\,\mu$m. In all measurements reported here we used
p-polarized fundamental and second harmonic light. To correct for a
varying laser intensity, a reference signal was used to record only
signals within an intensity interval of $\pm 5\%$. Measurements of the
linear reflectivity $R(\omega)$ were carried out with pump and probe
pulses focused to about $0.5\,$mm and $0.2\,$mm, respectively.
Frequency-doubled probe pulses were used to obtain $R(2\omega)$. For
linear reflectivity measurements the probe pulses were attenuated by a
factor of $5\times 10^{3}$. The measurements were carried out under
normal conditions at incident angles of $43^{\circ}$° (pump) and
$48^{\circ}$° (probe).  All data points in Figs.\ \ref{1}-\ref{4}
correspond to an average over about $500$ shots, and represent
relative changes of probe beam signals normalized to values at
negative delays.

\section{Data analysis}

Information about the electron temperature dependence of $\chi^{(2)}$
from time resolved SHG measurements on polycrystalline metals requires
 knowledge about (i) the magnitude of the three independent tensor
elements of $\chi^{(2)}$ and possible nonlocal contributions, (ii)
the variation of $\epsilon(\omega)$ and $\epsilon(2\omega)$ with
electron temperature $T_{e}$, and (iii) the electron temperature
relaxation.

\begin{figure}[htbp]
\epsfxsize=12.cm
\centerline{\epsffile{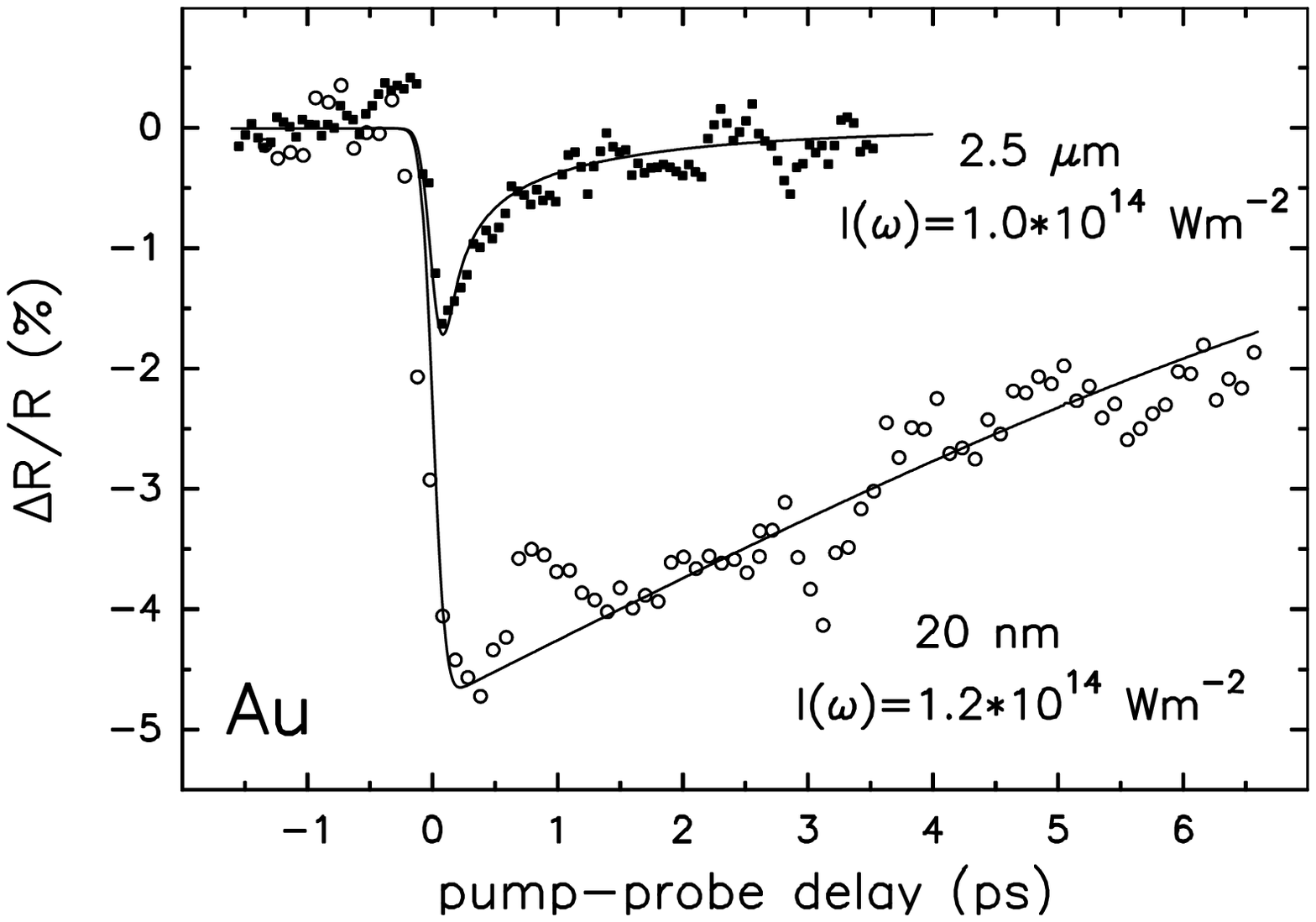}}
\vspace{2ex}
 \caption{\label{1}
Relative change of transient linear reflectivities at a photon energy of $2\,$eV for $2.5\,\mu$m (bulk) and $20\,$nm thick samples of
Au. Solid lines represent model calculations with the same
electron-phonon coupling in both cases.  Electron diffusion was taken
into account for the $2.5\,\mu$m sample but neglected for the $20\,$nm
film.  }
\end{figure}

To obtain the electron temperature dependence of $\epsilon(\omega)$
and $\epsilon(2\omega)$ and the time dependence of $T_{e}$, we
analyzed time-resolved measurements of linear reflectivities at photon
energies of fundamental ($2 \,$eV) and frequency-doubled light
($4\,$eV).  As an example, results of transient linear reflectivities
at $2\,$eV for Au (dots) of a $2.5 \, \mu$m (bulk) sample and a $20 \,
$nm thin film, for which electron diffusion is negligible, are
compared to corresponding model calculations (solid lines) in Fig.\ 
\ref{1}.
Although the ratio of the pump intensities used for the thin film and
the $2.5\,\mu$m sample was only $1.2 : 1$, the transient change of $R$
for the thin film is more than twice as large as that for thick metal.
The considerably longer relaxation time for the thin film also signals
reduced electron cooling and confirms the importance of diffusive heat
transport by electrons in bulk metals.

The model calculations  are
based on knowledge of the dielectric functions, $\epsilon(\omega)$ and
$\epsilon(2\omega)$, and of the electron temperature, $T_{e}$. The
calculational procedure followed here was described in \cite{Hoh_95}
and is as follows: (1) Tabulated experimental values of
$\epsilon(\omega)$ \cite{Pal_85} were fitted to the model of Rustagi
\cite{Rus_68} to derive the functional dependence of $\epsilon$ on
photon energy in the range $1-5\,$eV at room temperature. For the
calculation of the dielectric functions the approximation of electron
momentum independent transition matrix elements was used
\cite{Sun_94}.  (2) In the same model the electron temperature $T_{e}$
is contained in the Fermi-function. This allows to calculate
$\epsilon(T_{e})$ for both photon frequencies $\omega$ and $2\omega$
without any additional fit parameters. (3) Cooling of the electron
temperature with time was treated according to the two-temperature
model by Anisimov {\it et al.}\ \cite{Ani_75}. The use of the two-temperature 
model is justified by electron thermalization times of less than
$50\,$fs, predicted for our experimental conditions by the
Fermi-liquid theory \cite{Pin_66}. In this way we obtained
the time-dependence of $\epsilon(\omega)$ and $\epsilon(2\omega)$,
fitted to the data in Fig. \ref{1}.

No dependence of $R$ on pump-probe delay was found for Ag at $2\,$eV
and for Au at $4 \,$eV, while Cu showed a similar effect at both
photon energies.  Model calculations lead to good agreement with
transient reflectivity data for all three metals.

The variation of $\epsilon(\omega)$ and $\epsilon(2\omega)$ with
$T_{e}$ is now used to predict the effect of transient electron
temperatures on the Fresnel factors in SHG. Any deviation from
experimental data is then interpreted as electron temperature
dependence of the nonlinear susceptibility. Generally, this is
complicated by three independent tensor elements of $\chi^{(2)}$ as
well as possible nonlocal contributions.  For isotropic bulk material,
realized in polycrystalline metals, the {\it nonlocal response} generates
only P-polarized SHG. The corresponding field can be written in the
form (Eq. (22) in \cite{Sip_87}):
\begin{equation}
E_{P}^{nl}(2\omega)\propto \gamma \cdot [E_{s}^{2}(\omega)+E_{p}^{2}(\omega)]\, \, .
\end{equation}
{\it Dipole-allowed} surface contributions, on the other hand. are given by Eq. (30) in \cite{Sip_87}:
\begin{equation}
E_{S}(2\omega)\propto \chi_{xzx}\cdot E_{s}(\omega) E_{p}(\omega)\, \, ,
\end{equation}
\begin{eqnarray}
E_{P}(2\omega)&\propto& A \cdot \chi_{zxx} \cdot E_{s}^{2}(\omega)+ \nonumber \\
&&[B \cdot \chi_{zzz}+C \cdot \chi_{xzx}+D \cdot \chi_{zxx}]\cdot E_{p}^{2}(\omega) \, \, .
\end{eqnarray}
The different contributions can be separated by polarization dependent measurements.

When analyzing the SH polarization for different polarizations of the
fundamental, only P-polarized SHG was detectable for all three metals.
It varied with $\cos^{4}\Phi$, where $\Phi$ is the polarization angle
of the incident radiation.  Since no S-polarized SH radiation was
detectable for arbitrary input polarization, we conclude that
$\chi_{xzx}$ is vanishingly small. As discussed by Petrocelli {\it et
  al.}\ \cite{Pet_93} it is reasonable to assume $\chi_{xzx} \approx
\chi_{zxx}$ for polycrystalline metal surfaces. Hence, $\chi_{zxx}$ is
also small and $\gamma$ remains the only possible source of
P-polarized SH generated by s-polarized input fields. Since no SHG
yield was detected for this s-P polarization combination, $\gamma$ is
concluded to be negligible. Summarizing this discussion, the dipole
allowed $\chi_{zzz}$ connected to the near surface region where
symmetry is broken is the only source of SHG.

The second
harmonic amplitude can now be expressed in the form \cite{Sip_87}:
\begin{eqnarray}
\label{E2w}
E(2\omega,T_{e})\propto F(2\omega)\cdot\chi_{zzz}\cdot f(\omega)\cdot |E(\omega)|^{2}=&&\nonumber\\
{\cal T}_{p}(T_{e})
\epsilon(2\omega,T_{e}){\cal F}_{s}(T_{e})\cdot
\chi_{zzz}\cdot {\sl f}^{2}_{s}(T_{e}) {\sl t}^{2}_{p}(T_{e})\!\!\!\!\!\!&&\cdot |E(\omega)|^{2},
\end{eqnarray}

\noindent with 

\begin{eqnarray}
\label{eq_abb}
{\sl f}_{s}(T_{e})& =&
\frac{\sin{\Theta}}{\sqrt{\epsilon(\omega,T_{e})}}\,\, \,  , \,\, {\sl f}_{c}(T_{e})=\sqrt{1-{\sl f}_{s}(T_{e})^{2}}\, , \,\, \nonumber \\ 
{\sl t}_{p}(T_{e})&=&\frac{2\cos{\Theta}}{\sqrt{\epsilon(\omega,T_{e})}\,\cos{\Theta}+{\sl f}_{c}(T_{e})}\,  , 
\end{eqnarray}
where $\Theta$ is the angle of incidence.  Equivalent expressions for
${\cal F}_{s}$, ${\cal F}_{c}$, and ${\cal T}_{p}$ with
$\epsilon(2\omega,T_{e})$ are used, with lower case (capital) letters
denoting quantities of the fundamental (second harmonic) radiation.

The relative change of $|\chi_{zzz}|^{2}$ with $T_{e}$ can be
extracted from the differences $\Delta SH=SH(T_{e})-SH_{0}$ of
measured ($SH_{m}$) and predicted ($SH_{p}$) second-harmonic
intensities:
\begin{equation}
\label{chi2}
\frac{\Delta |\chi_{zzz}|^{2}}{|\chi_{zzz}|^{2}}=\frac{\Delta SH_{m}-\Delta SH_{p}}{SH_{p}(T_{e})}\, ,
\end{equation}
where $SH_{0}$ is the yield for the probe pulse alone.

\section{Second-harmonic reflectivities}

Results of pump-probe SHG measurements are shown in Fig.\ \ref{2} for
Cu and Ag, and in Fig.\ \ref{3} for Au.
 Plotted are the relative
changes of the probe SHG signal as a function of delay between pump
and probe pulses. The peaks around zero delay correspond to a coherent
artifact between pump and probe beams and will not be discussed here.
Predictions based on Eq.\ (\ref{E2w}), with constant $\chi^{(2)}$ and
the temperature dependence exclusively in the Fresnel factors, are
indicated by dotted lines. Solid lines represent smoothed relative
changes of $|\chi_{zzz}|^{2}$ with electron temperature, determined by
Eq.\ (\ref{chi2}).

 For Cu, the dielectric functions describe the
temperature dependence of the SHG data well, and we conclude that
$\chi^{(2)}$ is in first order independent of electron temperature in
the intensity range used. In contrast, for Ag and Au the temperature
dependence of SHG is almost entirely contained in the nonlinear
susceptibility.

\begin{figure}
\epsfxsize=12.cm
\vspace{2cm}
\centerline{\epsffile{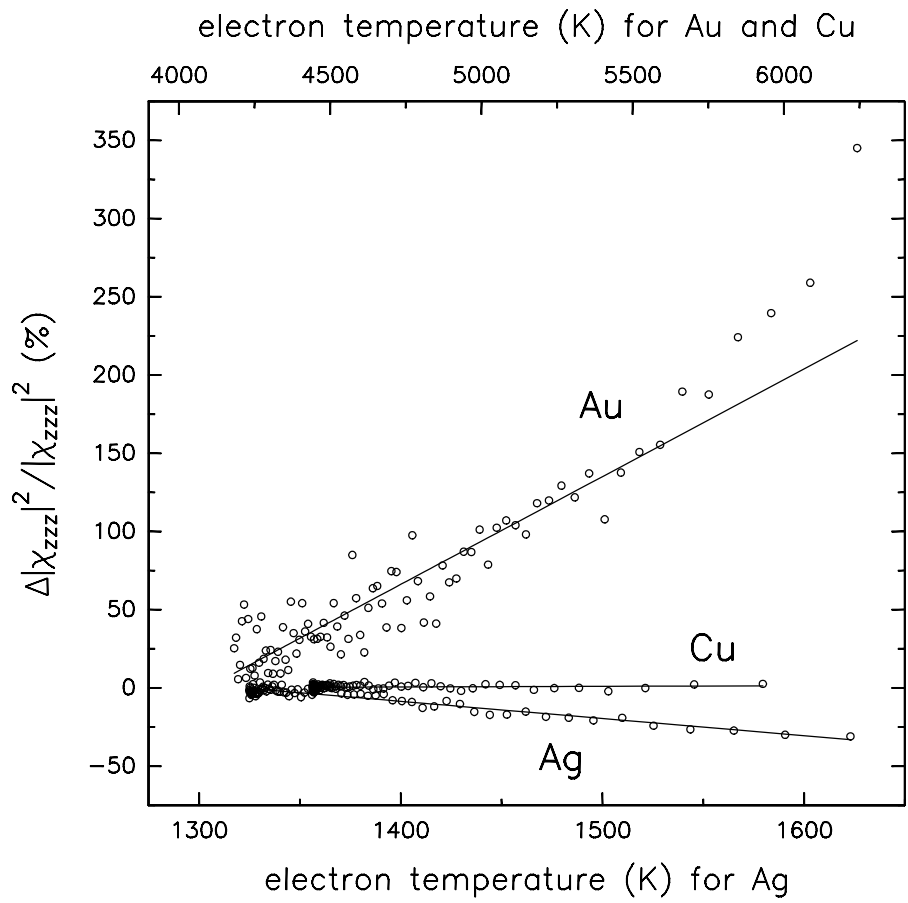}}
\vspace{2ex}
 \caption{\label{4}
  Relative changes of $|\chi_{zzz}|^{2}$ with electron temperature for Cu, Ag, and Au.}
\end{figure}

Another way of presenting the data derived from Eq.\ \ref{chi2} is
shown in Fig.\ \ref{4}, where the relative change of
$|\chi_{zzz}|^{2}$ is plotted versus electron temperature. 

For this, we used the correlation between electron temperature and
time scale given by the analysis of the linear reflectivities (Fig.\ 
\ref{1}). From Fig.\ \ref{4} it is obvious that for Cu $\chi_{zzz}$
does not depend on temperature. For Ag and Au, $\Delta
|\chi_{zzz}|^{2}$ depends linearly on $T_{e}$. The different signs of
the slopes reflect the mismatch between the photon energy at $\omega$
(Au) and $2\omega$ (Ag) and the interband transition thresholds.
Although no theoretical prediction about the temperature dependence of
$|\chi_{zzz}|^{2}$ is reported in the literature, it is reasonable to
assume that $|\chi_{zzz}|^{2}$, like the linear reflectivity, depends
linearly on small temperature variations. Thus, the linear
dependencies of $|\chi_{zzz}|^{2}$ on $T_{e}$ for Ag and Au indicate
similar electron dynamics at the surface and in the bulk. Deviations
for Au above $5500\,$K may actually originate from different electron
dynamics at the surface and in the bulk during the first $200\,$fs but
may also reflect a nonlinear temperature dependence of
$|\chi_{zzz}|^{2}$.

Regarding surface contaminations, no evidence for it was found for Ag
and Au. On copper surfaces an oxide layer reduces the total SHG yield
\cite{BBD_91,ToA_86} and may in fact cause that $|\chi_{zzz}|^{2}$ is
independent of electron temperature. Therefore, the measurements on
copper should be repeated under UHV-conditions.  Nevertheless, the
influence of the oxide layer on the electron temperature dependence of
the Fresnel factors remains small, as determined by the time
dependence of the linear reflectivity.  Thus, notwithstanding whether or
not $\chi^{(2)}$ is influenced by the copper oxide/copper interface,
we have demonstrated that there are systems like Au and Ag, where time
resolved SHG monitors the electron temperature relaxation at the
surface, and others, like Cu in air, where SHG is only sensitive to
the electron temperature in the bulk.

\section{Conclusion}

Femtosecond time-resolved SHG with photon energies near the interband
transition threshold was demonstrated to be a powerful tool for
investigating electron dynamics in noble metals. Transient linear
reflectivities were combined with pump-probe SHG data to separate the
dependence of Fresnel factors and $\chi^{(2)}$ on electron
temperature. A contrasting behavior was found. In case of Cu,
transient electron temperatures affect the Fresnel factors,
$\chi^{(2)}$ is independent of electron temperature, and there is no
surface sensitivity. For Ag and Au, $\chi^{(2)}$ varies with electron
temperature and pump-probe SHG monitors the electron relaxation near
the surface.

{\it Acknowledgements.}
  This work was supported by the Deutsche Forschungsgemeinschaft,
  Sonderforschungsbereich 290. The experiments were performed at the
  Laboratory for Applications of the Max-Born-Institut Berlin, and the
  cooperation of Dr.\ F. Noack and his team is gratefully
  acknowledged.  We also thank Dr.\ W. H{\"u}bner for fruitful
  discussions.
%
%
%
%
\end{document}